\documentclass[aps,reprint]{revtex4-1}
\usepackage[utf8]{inputenc}
\usepackage[english]{babel}
\usepackage{amsmath}
\usepackage{amssymb}
\usepackage{float}
\usepackage{graphicx}
\usepackage{hyperref}

\begin{document}

\title{\textbf{Electron-irradiation effects on monolayer MoS$_2$ at elevated temperatures}}
\author{Carsten Speckmann$^{1,2,*}$, Kimmo Mustonen$^{1}$, Diana Propst$^{1,2}$,\\ Clemens Mangler$^1$, and Jani Kotakoski$^{1,*}$\\
$^1$University of Vienna, Faculty of Physics, Boltzmanngasse 5,\\ 1090 Vienna, Austria\\
$^2$University of Vienna, Vienna Doctoral School in Physics, Boltzmanngasse 5,\\ 1090 Vienna, Austria\\
$^*$Email: carsten.speckmann@univie.ac.at and jani.kotakoski@univie.ac.at}
\date{\today}


\begin{abstract}

    The effect of electron irradiation on 2D materials is an important topic, both for the correct interpretation of electron microscopy experiments and for possible applications in electron lithography.
    After the importance of including inelastic scattering damage in theoretical models describing beam damage and the lack of oxygen-sensitivity under electron irradiation in 2D MoS$_2$ were recently shown, the role of temperature has remained unexplored on a quantitative level.
    Here we show the effect of temperature on both the creation of individual defects as well as the effect of temperature on defect dynamics.
    Based on the measured displacement cross section of sulphur atoms in MoS$_2$ by atomic resolution scanning transmission electron microscopy, we find an increased probability for defect creation for temperatures up to $150^\circ$C, in accordance with theoretical predictions.
    However, higher temperatures lead to a decrease of the observed cross sections.
    Despite this apparent decrease, we find that the elevated temperature does not mitigate the creation of defects as this observation would suggest, but rather hides the created damage due to rapid thermal diffusion of the created vacancies before their detection, leading to the formation of vacancy lines and pores outside the measurement's field of view.
    Using the experimental data in combination with previously reported theoretical models for the displacement cross section, we estimate the migration energy barrier of sulphur vacancies in MoS$_2$ to be $0.26 \pm 0.13$~eV.
    	These results mark another step towards the complete understanding of electron beam damage in MoS$_2$.

\end{abstract} 

\maketitle
\newpage

\section{Introduction}

	Understanding electron beam-induced material changes during (scanning) transmission electron microscopy [(S)TEM] at the atomic level would allow distinguishing between observed intrinsic effects from those caused by the imaging process.
	These insights could then further be utilized for example in electron beam lithography~\cite{pease_electron_1981, groves_electron_2014, chen_nanofabrication_2015}.
	Notably, although electron beam damage has been investigated for decades~\cite{egerton_radiation_2004, egerton_control_2013}, a detailed quantification has only been possible since the advent of two-dimensional (2D) materials~\cite{susi_quantifying_2019}.
    Electron beam damage has already been studied for a variety of 2D materials, including graphene~\cite{meyer_accurate_2012,meyer_erratum_2013,susi_isotope_2016}, hexagonal boron nitride~\cite{meyer_selective_2009,jin_fabrication_2009,kotakoski_electron_2010, bui_creation_2023}, MoS$_2$~\cite{komsa_two-dimensional_2012,kretschmer_formation_2020, speckmann_combined_2023}, MoSe$_2$~\cite{lehtinen_atomic_2015, lin_vacancy-induced_2015}, MoTe$_2$~\cite{elibol_atomic_2018,ahlgren_atomic-scale_2022} and phosphorene~\cite{yao_situ_2020, speckmann_electron-beam-induced_2024}.
	However, only recently the role of inelastic scattering was included into theoretical models for electron beam damage in non-conducting 2D materials~\cite{kretschmer_formation_2020, speckmann_combined_2023, yoshimura_quantum_2023}, while previous reports only tried to estimate its importance indirectly~\cite{algara-siller_pristine_2013, lehnert_electron_2017}.
	Following these reports, other studies explored the effect adatoms have on the displacement process~\cite{jain_adatom-mediated_2024} and elucidated the differences in cross section values of different layers in multilayer structures~\cite{quincke_defect_2024}.
    An additional complication is that in some cases the non-ideal vacuum within the microscope can result in chemical etching also changing the atomic structure of the sample~\cite{leuthner_scanning_2019}.
	
	Interestingly, for MoS$_2$, imaging above temperatures of $400^\circ$C has been reported to reduce the number of created defects~\cite{lin_atomic_2014}.
	However, when they do appear, sulphur vacancies agglomerate in vacancy lines, rather than staying as single or double vacancies~\cite{lin_vacancy-induced_2015, chen_ultralong_2018, chen_situ_2019, inani_step-by-step_2021, li_growth_2024}.
	At higher temperatures, also the formation of pores has been reported~\cite{chen_ultralong_2018, chen_situ_2019, inani_step-by-step_2021}.
    Luckily, in the case of MoS$_2$, chemical reasons behind the pore formation have been ruled out~\cite{ahlgren_atomic-scale_2022}, which simplifies the interpretation of the results.
	Indeed, agglomeration of vacancies has been attributed to the high mobility of sulphur vacancies~\cite{chen_atomically_2017, chen_ultralong_2018, chen_situ_2019} and it has also been suggested to allow some vacancies to escape the imaged field of view (FOV)~\cite{chen_atomically_2017, chen_ultralong_2018}.
	Even grain boundaries and vacancy lines have been reported to be mobile during imaging~\cite{chen_situ_2019}. 
    However, the migration barrier for sulphur vacancies has been estimated to be fairly high (ca. $2.3$~eV for an isolated vacancy~\cite{komsa_point_2013, komsa_native_2015}), but to decrease significantly (down to $0.8$~eV~\cite{komsa_point_2013, li_growth_2024}) when other vacancies are nearby.
	Experimentally, a barrier based on the migration of domain boundaries has been estimated to be even lower, only $0.6$~eV~\cite{precner_evolution_2018}.
	However, a systematic and quantitative study on the effect of temperature on the creation and migration of defects under electron irradiation has not been carried out yet.	
	At room temperature, it has been shown~\cite{kretschmer_formation_2020,yoshimura_quantum_2023,speckmann_combined_2023} that it is necessary to include inelastic scattering effects to adequately describe the displacement process.
    In our previously proposed model~\cite{speckmann_combined_2023}, the experimentally obtained displacement cross section values can be described satisfactorily with an impact ionization model~\cite{kretschmer_formation_2020} which results in excitation lifetimes comparable with previous optical measurements~\cite{korn_low-temperature_2011, lin_physical_2017, palummo_exciton_2015}.
    The model predicts an increasing cross section at elevated temperatures, arising from the enhanced contribution of lattice vibrations~\cite{susi_quantifying_2019}.
    
    In this work, we measure the displacement cross section of S atoms in MoS$_2$ at temperatures up to $550^\circ$C using scanning transmission electron microscopy at acceleration voltages of $60$ and $90$~kV.
	At a temperature of $150^\circ$C, the displacement cross section increased compared to its room temperature value~\cite{speckmann_combined_2023}, which is in agreement with theoretical models attributing this to a thermal activation of phonons, increasing the maximum transferred energy of the beam electrons to the material~\cite{meyer_accurate_2012, meyer_erratum_2013, susi_isotope_2016, chirita_three-dimensional_2022}.
	However, at higher temperatures the measured cross section decreases significantly.
	We attribute the apparent decrease to thermal diffusion of sulphur vacancies before they can be detected with the microscope.
	We use the difference in experimental and theoretical cross sections to estimate the migration energy barrier of sulphur vacancies in MoS$_2$ to be $0.26 \pm 0.13$~eV, which is within the same order of magnitude as previous theoretical calculations~\cite{komsa_point_2013, li_growth_2024} and experimental estimates~\cite{precner_evolution_2018}.
	Overall, our results show that elevated temperatures do not mitigate defect creation by electron irradiation, but only obscure the effects of the electrons impinging on the sample due to thermal diffusion.
	These results mark another step towards the complete understanding of electron beam damage in MoS$_2$, which is important for any future device fabrication.

\section{Results and Discussion}
	For this work, MoS$_2$ samples were grown via chemical vapour deposition (CVD) and were subsequently transferred onto Fusion AX heating chips from Protochips, resulting in a free-standing MoS$_2$ layer (see the Methods section for more details).
	To distinguish the effects of the increased temperatures on both the elastic and inelastic scattering damage, two electron energies were used, $60$~keV for the mostly inelastic scattering damage regime and $90$~keV for the pure elastic scattering damage regime.
	Temperatures used during the course of this study are $150^\circ$C, $300^\circ$C, $450^\circ$C, $550^\circ$C and $750^\circ$C and a heating rate of $5^\circ$C/s was used for all experiments. 
	Note that no data could be recorded at $750^\circ$C as the sample decomposed at this temperature before any measurements could be conducted.
	This is in good agreement with a previous report identifying the decomposition of MoS$_2$ in ultra-high vacuum (UHV) to take place between $700^\circ$C and $800^\circ$C~\cite{chen_thermal_2019}.
		
\begin{figure}[ht!]
\centering
\includegraphics[width=0.45\textwidth]{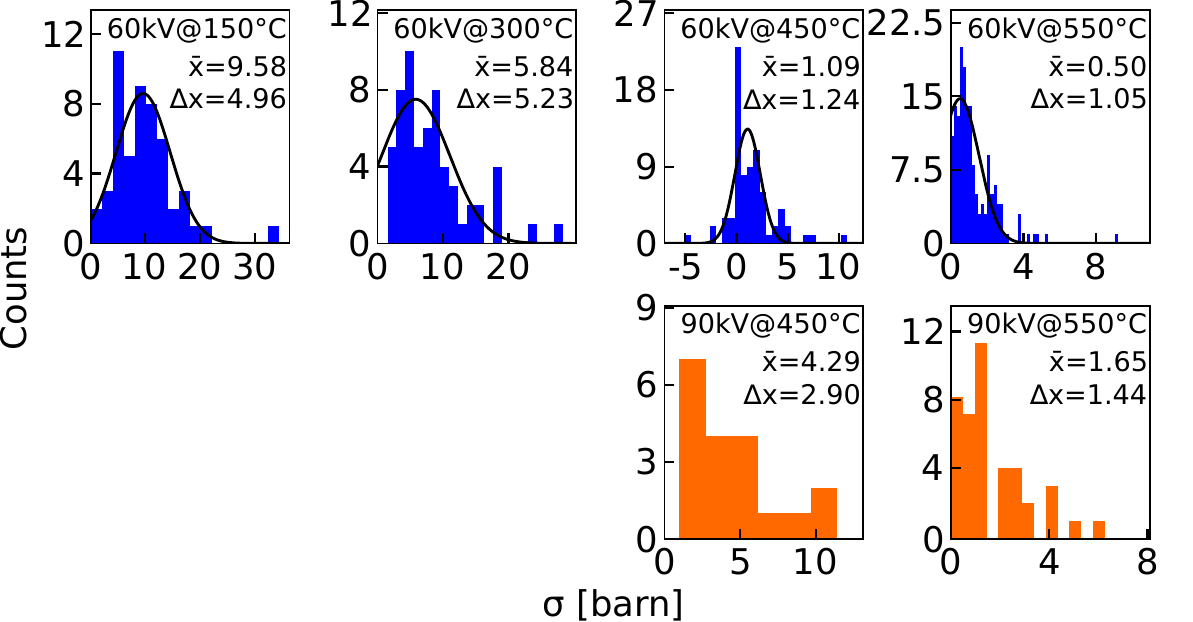}
\caption{{\bf Statistics of sulphur vacancy formation at elevated temperatures.}
    Histograms show the statistics of the measured displacement cross sections for different electron energies and temperatures.
    The top row shows experiments carried out with an electron energy of $60$~keV, while the bottom row shows results from experiments carried out at $90$~keV.
    Voltages and temperatures are stated on the top of each histogram.
    Here, one count represents the results of a single image series.
    The solid lines correspond to fits of a normal distribution to the datasets recorded at $60$~keV with means $\bar{x}$ and standard deviations $\Delta x$ of each fit given in the top-right corner of each plot underneath the temperatures.
    The values stated for the datasets recorded at $90$~keV correspond to the mean and standard deviation of the individual values.}
\label{fig:DatasetsMoS2_highT}
\end{figure}

	Measurements on the displacement cross section of MoS$_2$ were conducted in the same way as reported in our previous study~\cite{speckmann_combined_2023}, resulting in similar image series of high-angle annular dark field (HAADF) images.
	In each of these image series, sulphur vacancies were identified and the number of created vacancies per frame was calculated.
	This was then correlated with the beam current to calculate an average value for the created vacancies per electron impinging on the sample, $\bar{N}$, which can be converted into the displacement cross section, $\sigma = \bar{N}/\rho$ (see Methods section for more details).
	Here, $\rho$ indicates the areal density of the sulphur sites in MoS$_2$, which in this case is identical to the inverse unit cell area, calculated using a lattice constant of $3.19$~{\AA}~\cite{ahmad_comparative_2014}, as a single vacancy was assumed to have been created first whenever a double vacancy was observed.
	The resulting statistics of this analysis for each set of measurements are shown in Fig.~\ref{fig:DatasetsMoS2_highT} with the top row depicting the measurements at $60$~keV and the bottom row showing the measurements at $90$~keV. 
	Here, each count corresponds to the calculated displacement cross section based on a single image series.
	Note that each image series contains multiple images recorded within the same sample area, and the number of vacancies created by the electron beam was measured as a function of the electron dose.
	As the already present number of defects in the first frame was used as a baseline, negative cross sections might appear in case vacancy healing or vacancy migration takes place after these defects were identified in a previous frame.
	Furthermore, although each individual displacement process follows a Poissonian distribution~\cite{susi_silicon--carbon_2014}, we record here several of these events per frame with multiple frames per image series, and therefore effectively measure an average over a number of these processes.
	Thus, according to the central limit theorem~\cite{polya_uber_1920}, these follow a normal distribution.	
	The temperature $T$ at which the respective measurements were carried out is indicated in the top right corner of each histogram.
	A Gaussian distribution was fitted to the shown statistics in the top row with the means $\bar{x}$ and standard deviations $\Delta x$ visible underneath the temperatures.
	No Gaussian distribution could be fitted to the data acquired at $90$~keV, thus, the mean of the individual values and its standard deviation are stated in the histograms instead.
	Note that no liable data could be recorded at an electron energy of $90$~keV below temperatures of $450^\circ$C as a pore would always emerge within the field of view after just a single frame, making it impossible to estimate the number of missing atoms. 	
	
\begin{figure}[ht!]
\centering
\includegraphics[width=0.45\textwidth]{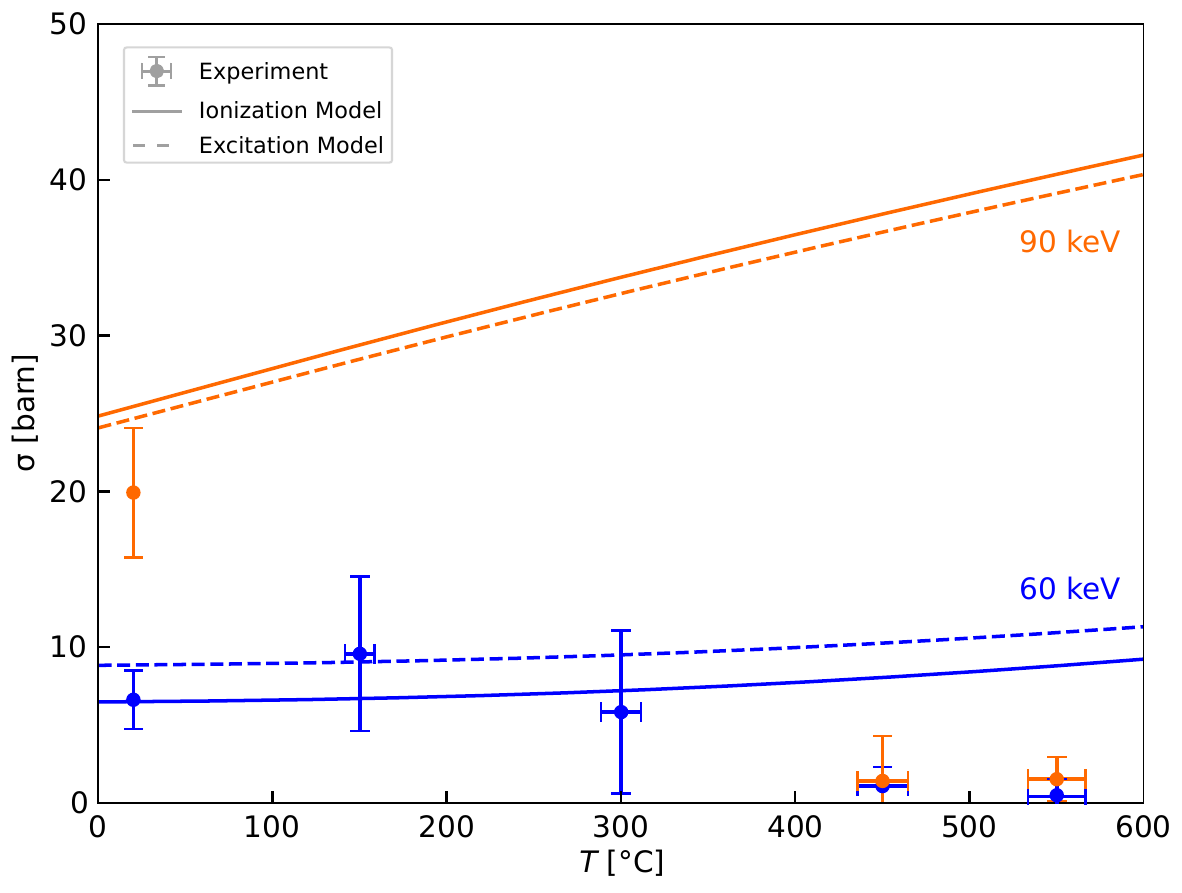}
\caption{{\bf Experimental displacement cross section values as a function of temperature.}
    Total displacement cross section values for single sulphur atoms in MoS$_2$ as a function of temperature measured with electron energies of $60$~keV (filled blue circles) and $90$~keV (filled orange circles).
    Room temperature values are taken from Ref.~\cite{speckmann_combined_2023}.
    Lines correspond to theoretical models~\cite{speckmann_combined_2023}.
    The solid lines indicate the impact ionization model with one excited state and the dashed lines indicate the electronic excitation model with two excited states (see Ref.~\cite{speckmann_combined_2023} for further details) using the fit parameters stated in Table~I of Ref.~\cite{speckmann_combined_2023}.
    Blue lines correspond to an electron energy of $60$~keV, while orange lines correspond to an energy of $90$~keV.}
\label{fig:Fits_highT}
\end{figure}	

	The cross section values as well as two theoretical models are plotted as a function of temperature in Fig.~\ref{fig:Fits_highT}.
	The models shown here were reported in our previous study	~\cite{speckmann_combined_2023}, including both elastic and inelastic scattering damage for the explanation of beam damage in MoS$_2$ with two distinct ways of incorporating the inelastic scattering processes.
	The ionization model describes the inelastic scattering as electron impact ionization of sulphur atoms, while the excitation model describes the inelastic contributions as valence excitation instead.
	Both models explain the experimentally observed datapoints reasonably well at room temperature, with only the ionization model resulting in a quantitatively good agreement with previous reports on the exciton lifetimes in MoS$_2$~\cite{speckmann_combined_2023}.
	
\begin{figure}[ht!]
\centering
\includegraphics[width=0.45\textwidth]{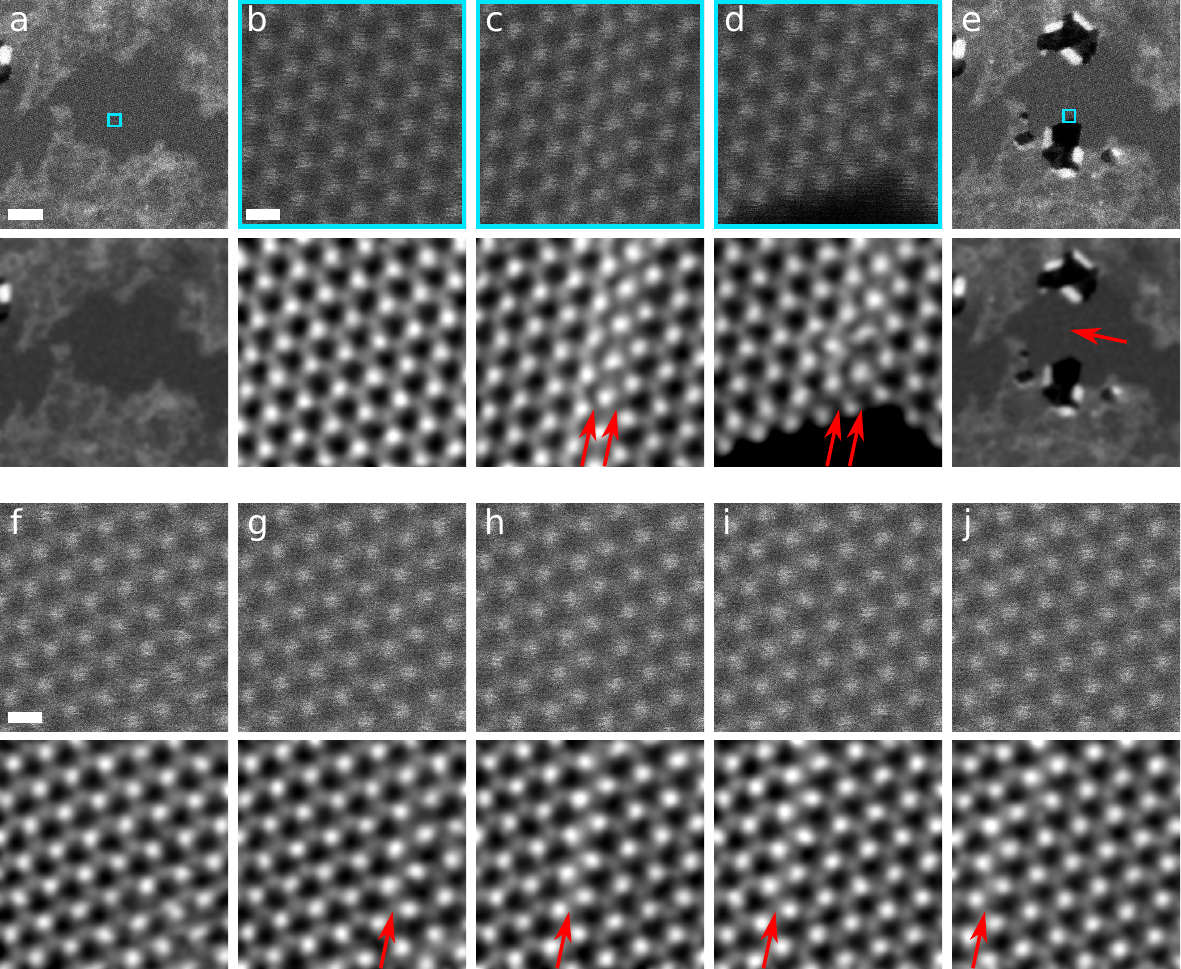}
\caption{{\bf Example STEM-HAADF image series of MoS$_2$ at elevated temperatures and the migration of vacancy lines.}
    Panels (a-e) show five images of an image series where each frame is a $512~\mathrm{px}\times 512~\mathrm{px}$ image recorded within a frame time of $0.88$~s.
    Images were taken at an acceleration voltage of $60$~kV and at a temperature of $550^\circ$C.
    Images in the top row are as-recorded, whereas a Gaussian blur ($4$~px in panels a and e and $8$~px in panels b-d) has been applied to the images in the second row.
    Panels (f-j) show five subsequent images of another image series recorded using the same parameters but at $450^\circ$C.
    Images in the top row are as-recorded, whereas a Gaussian blur of $8$~px has been applied on the images in the second row.
    For HAADF images, the intensity increases with the atomic number $Z$ of the atoms located at the imaged position~\cite{pennycook_z-contrast_1989,pennycook_high-resolution_1991,hartel_conditions_1996} proportional to $Z^{1.64}$ for the HAADF detector used here~\cite{krivanek_atom-by-atom_2010}.
    Thus, Mo atoms are the brightest spots, S dimers are less bright, and single S vacancies are again a bit less bright than the S dimers.
    Red arrows are used to indicate vacancy lines in the MoS$_2$ lattice and cyan rectangles in panels a and e indicate the positions of the zoomed in images in panels (b-d).
    The scale bars are $5$~nm for images in panels a and e and $3$~\r{A} for the other images.}
\label{fig:highTseries}
\end{figure}

	In Fig.~\ref{fig:Fits_highT}, the experimental values measured at $60$~keV (blue) start to increase with increasing temperature, following the predictions of the models.
	This increase can be described by the thermal activation of phonons that influence the maximum transferable energy of the electron to the S atom~\cite{meyer_accurate_2012, meyer_erratum_2013, susi_isotope_2016, chirita_three-dimensional_2022}.
	A similar trend is also expected for $90$~keV electrons, which could however not be measured due to the immediate appearance of a pore, rather than individual sulphur vacancies.
	However, a decrease in cross section can be observed at temperatures above $150^\circ$C, completely deviating from the theoretical models above $300^\circ$C.
	This deviation is especially pronounced for an electron energy of $90$~keV (orange) at temperatures $\geq 450^\circ$C.
	A possible explanation for the decrease of the apparent cross section is the filling of vacancies due to migrating atoms on the sample surface~\cite{inani_step-by-step_2021}, similar to graphene~\cite{postl_indirect_2022}.
	Another possibility would be a faster recombination of excitations caused by inelastic scattering events at elevated temperatures as previous reports suggest that the presence of electronic excitations in MoS$_2$ can lead to a significant reduction of the displacement threshold energy~\cite{kretschmer_formation_2020, speckmann_combined_2023, yoshimura_quantum_2023}.
	This would however not explain a similar reduction of the cross section within the purely elastic scattering regime ($90$~keV).
	However, the most prominent cause for the apparent reduction of the cross section seems to be thermal diffusion~\cite{chen_atomically_2017, chen_ultralong_2018}, as shown in Fig.~\ref{fig:highTseries}.

	Fig.~\ref{fig:highTseries}(a-e) show an example image series recorded at an acceleration voltage of $60$~kV and at a temperature of $550^\circ$C.
	Fig.~\ref{fig:highTseries}a shows a clean patch of MoS$_2$ enclosed by contamination, while a small pore including a metal cluster is partially visible at the top left corner of the image.
	The following three images were recorded at the position marked with the cyan squares over a period of $145$~s between panels b and d.
	Note that no individual sulphur vacancies can be seen in these images, as was the case for most images taken at $550^\circ$C. 
	Rather than individual vacancies, vacancy lines of various sizes could be observed along the zig-zag (ZZ) directions, highlighted by the red arrows.
	Here, vacancy lines refer to several sulphur vacancies agglomerating along the zig-zag direction, resulting in a locally sulphur depleted phase.  
	Fig.~\ref{fig:highTseries}d then shows that the vacancy line ends with a pore that expands into the FOV over time.
	A similar defect creation behaviour at elevated temperatures was reported for graphene, where a pore would only form after extensive vacancy lines were already created outside the scanned FOV~\cite{dyck_role_2023}.
	The amount of damage created during the irradiation with the electron beam is shown in Fig.~\ref{fig:highTseries}e, where several pores can be seen to have emerged outside the FOV of the zoomed-in images recorded in between panels a and e with bright clusters (likely Mo) at their edges.
	Observations of pores outside the FOV, together with the absence of individual vacancies in many images recorded at $450^\circ$C and $550^\circ$C indicate that most vacancies created by the electron beam migrate out of the FOV before they can be detected, forming pores and/or vacancy lines.
	This rapid thermal diffusion might be the most prominent reason for the previously observed decrease of atomic displacements when combining electron irradiation with heating above temperatures of $400^\circ$C~\cite{lin_atomic_2014}.
	Thus, elevated temperatures only obscure the creation of vacancies by electron irradiation, but do not mitigate it.
	Note that all pores that could be observed during these measurements were (partially) located within the contaminated part of the sample, speaking for the pinning of defects by the contamination.
	This behaviour might be explained by the fact that sulphur vacancies in MoS$_2$ have dangling bonds, and are therefore likely to bond to contamination once in contact with it.
	A similar behaviour has been reported in graphene, where contamination was found to hide defects by accumulating on top of it~\cite{leuthner_scanning_2019}.	
	
	The vacancy lines are highly mobile as can be seen in Fig.~\ref{fig:highTseries}(f-j).
	Fig.~\ref{fig:highTseries}f shows an almost perfect hexagonal MoS$_2$ lattice.
	In the subsequent images, with a frame time of $0.88$~s, a vacancy line can be seen to have migrated into the FOV from outside (Fig.~\ref{fig:highTseries}g), migrating from right to left.
	The vacancy line is highlighted with a red arrow.
	During this process, a transitional state of the migration process can be seen in Fig.~\ref{fig:highTseries}j in which the vacancy line is spread across two sulphur rows.
	These line defects can form if sulphur vacancies accumulate along a line, and its width is determined by the number of lines where vacancies are present~\cite{komsa_point_2013, han_stabilities_2015, chen_ultralong_2018, wang_detailed_2016}.	
	Our observation seem to agree with previous theoretical and experimental studies on vacancy lines in MoS$_2$.
	They have found a preference for the formation along the ZZ direction~\cite{komsa_point_2013, han_stabilities_2015, chen_ultralong_2018, chen_situ_2019} with pores and Mo clusters forming at their ends~\cite{chen_ultralong_2018, chen_situ_2019}. 
	Furthermore, it has been predicted that vacancy lines can act as channels for the migration of S and Mo atoms towards the forming pore~\cite{chen_ultralong_2018} due to a reduction of the diffusion energy barrier~\cite{chen_ultralong_2018, li_growth_2024}.
	Vacancy lines have also been predicted to locally decrease the electronic band gap of MoS$_2$ depending on the width of the vacancy line~\cite{han_stabilities_2015, wang_detailed_2016, gao_defect_2021} up to the point where the vacancy line becomes metallic once four rows of sulphur vacancies are present~\cite{wang_detailed_2016}, opening the possibility for defect-engineered devices.
	
	Based on the above presented observations, the difference between the theoretically predicted and experimentally observed cross sections is likely to be primarily caused by migration.
	Thus, this difference can be used to estimate the migration barrier of sulphur vacancies in MoS$_2$, which was previously only estimated by theoretical means~\cite{komsa_point_2013, komsa_native_2015, li_growth_2024} and by an indirect experimental estimate based on the migration of domain boundaries~\cite{precner_evolution_2018}. 
	As also discussed in our previous study on the migration of carbon adatoms on graphene~\cite{postl_indirect_2022} the number of migration steps $\mu$ the defect taken during one image at a given temperature $T$ is
\begin{equation}
	\mu(T) = f_0 t_F \exp{\left(\frac{-E_m}{k_B T}\right)},
\label{eq:jumps}
\end{equation}	
\noindent
	where $f_0$ is the attempted jump frequency, $t_F$ the frame time, $E_m$ the migration energy barrier and $k_B$ the Boltzmann constant.
	When a vacancy is created within the MoS$_2$ structure, a certain time $t_d$ passes until this defect can be detected via the electron beam. 
	The specific time until a defect is detected is determined by the position of said defect in relation to the position of the electron beam and its scan parameters.
	Without thermal diffusion, each defect will be detected once the electron beam scans across its position, which is assumed for the theoretical models.
	However, if the vacancy is able to jump due to thermal activation, its position is constantly changing, possibly resulting in the vacancy migrating to a location outside the scanned area before it can be detected.
    The probability, $p$, for this to happen is directly proportional to $\mu(T)$.
    Thus, the number of observed defects, $N^\prime$, can be written as
\begin{equation}
	N^\prime = N(1-p),
\label{eq:approx}
\end{equation}    
    with the total number of created defects $N$.
    As a result, the ratio of the observed cross section $\sigma_\mathrm{exp}$ to the theoretical cross section $\sigma_\mathrm{theo}$ can be written as
\begin{equation}
	1-\frac{\sigma_\mathrm{exp}}{\sigma_\mathrm{theo}} \propto \exp{\left(\frac{-E_m}{k_B T}\right)}.
\label{eq:approx}
\end{equation}	

\begin{figure}[ht!]
\centering
\includegraphics[width=0.45\textwidth]{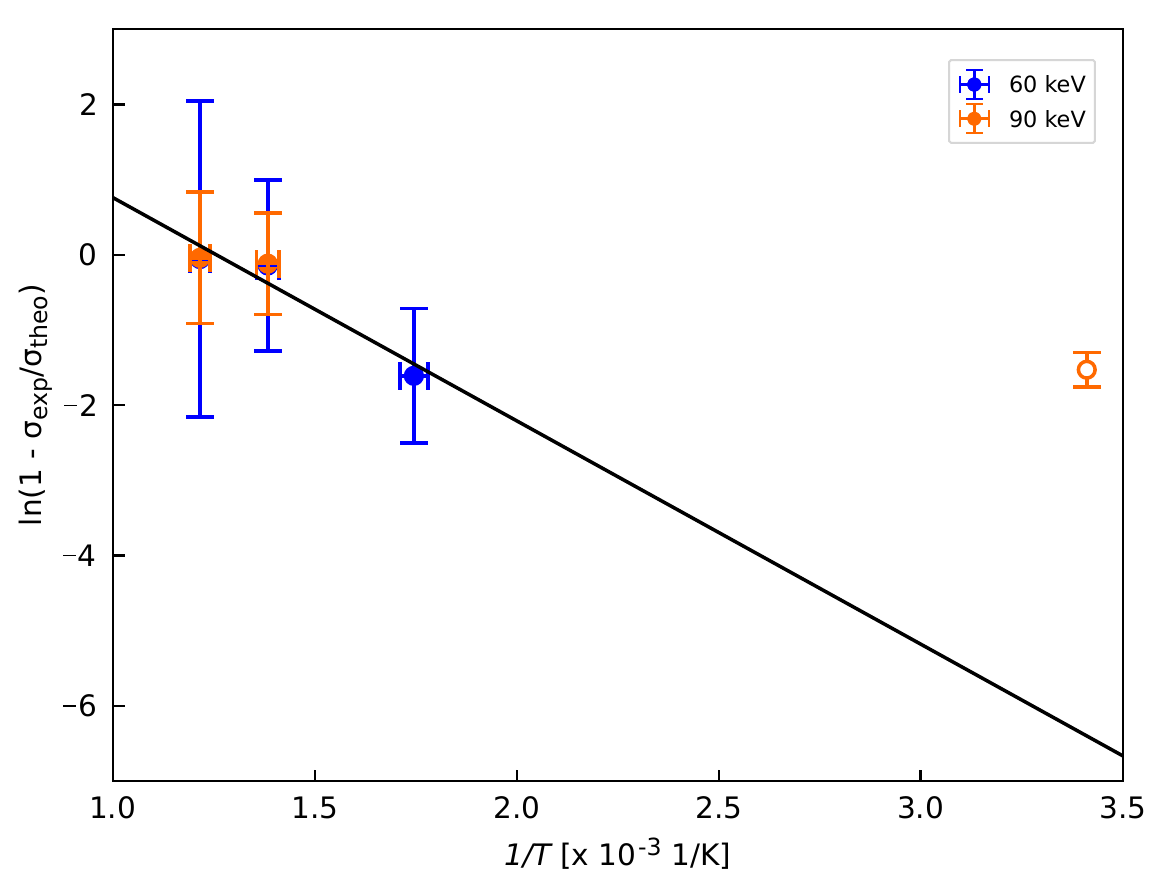}
\caption{{\bf Ratio of the experimental and theoretical displacement cross section as a function of temperature.}
    The natural logarithm of $1$ minus the experimentally observed displacement cross sections for single sulphur atoms in MoS$_2$ normalized to their respective theoretical predictions is plotted as a function of one over temperature while a linear regression, shown in black, is fitted to the datapoints.
    Blue and orange datapoints correspond to measurements done at electron energies of $60$ and $90$~keV, respectively.
    Additionally, the room temperature value measured at $90$~keV is shown with an open orange circle}
\label{fig:Fit_Em}
\end{figure}	

    The corresponding Arrhenius plot is shown in Fig.~\ref{fig:Fit_Em} where a linear regression (black line) is fitted to $1 - \sigma_\mathrm{exp}/\sigma_\mathrm{theo}$ (blue and orange circles).
    This plot also shows the room temperature value at $90$~keV (open circle) for reference as this value is also below the theoretical predictions even within it uncertainties (see Fig.~\ref{fig:Fits_highT}).
    Room temperature values at $60$~keV could not be included in this plot, as they exceed the theoretical predictions, thus, resulting in undefined values when using the natural logarithm.
    The likely reason for the lowered experimental cross section at room temperature for $90$~keV is the fact that measurements were mostly stopped after a single frame due to the rapid damaging of the material, thus, limiting the statistical significance.  
	As the migration of vacancies should only affect the observed cross section values once the observed experimental cross sections are decreasing again, only values recorded above a temperature of $150^\circ$C were included in the fit.
	Additionally, including the room temperature value at $90$~keV would result in an unphysically low migration barrier.
	
	In both cases, the fit values at room temperature corresponds to a $\sigma_\mathrm{exp}/\sigma_\mathrm{theo}$ ratio of over $99.5$~\%, validating the assumption that practically all defects are observed at that temperature. 
 	The fit results in a migration energy barrier of $0.26 \pm 0.05$~eV when using the ionization model and $0.15 \pm 0.03$~eV when using the excitation model as a theoretical reference.
	However, the actual uncertainty should be significantly higher due to the uncertainty of the individual datapoints.
	We estimate the maximum reasonable uncertainty of the migration barrier to be $\sim 50$~\% by fitting linear regressions though all datapoints, minimizing and maximizing the slopes, respectively.
	This results in values of $0.26 \pm 0.13$~eV and $0.15 \pm 0.08$~eV for the ionization and excitation model, respectively.
	These values are within the same order of magnitude compared to the theoretical migration barrier of vacancies next to other vacancies of $0.8$~eV~\cite{komsa_point_2013, li_growth_2024} and previous experimental estimations based of the migration of domain boundaries of $0.6$~eV~\cite{precner_evolution_2018}.
	However, the value obtained by the excitation model seems to be too low for the observed effects.
	This may indicate that the ionization model is better suited for the estimation of the vacancy migration barrier, which, in addition to the fact that it gives more reasonable values for excitation lifetimes~\cite{speckmann_combined_2023}, suggests that this picture might be better suited to explain inelastic scattering effects in MoS$_2$ in general.
	As the defects observed during the course of these measurements mostly appear next to each other, it is reasonable to assume that the migration barrier estimated through these measurements is not the pristine energy barrier, but rather corresponds to a defective structure.

\section{Conclusions}

	In conclusion, our measurements show the effect of elevated temperatures on the observed sulphur displacement cross sections in monolayer MoS$_2$.
	While temperatures of up to $150^\circ$C result in an increase of the cross section in agreement with theoretical models~\cite{meyer_accurate_2012, meyer_erratum_2013, susi_isotope_2016, chirita_three-dimensional_2022}, higher temperatures cause a stark decrease in the cross section, deviating from the model predictions.
	We show that this decrease originates from thermal diffusion of sulphur vacancies out of the field of view before they can be detected.
	The difference in experimental and theoretical cross sections is used to estimate the migration energy barrier of sulphur vacancies in MoS$_2$, resulting in a value of $0.26 \pm 0.13$~eV, which is within the same order of magnitude as previous theoretical calculations~\cite{komsa_point_2013, li_growth_2024} and experimental estimates~\cite{precner_evolution_2018}.
	Thus, our results show that elevated temperatures do not mitigate defect creation by electron irradiation, but only obscure the effects of the electrons impinging on the sample.
	These results mark another step towards the complete understanding of electron beam damage in MoS$_2$.

\section*{Methods}

    {\bf Sample preparation} 
    The MoS$_2$ sample was grown at Trinity College Dublin using CVD on a SiO$_2$ substrate following the recipe described in their previous study~\cite{obrien_transition_2014}.
    The MoS$_2$ was afterwards transferred in air onto a golden transmission electron microscopy grid covered with a holey membrane of amorphous carbon (Quantifoil R 1.2/1.3 Au grid) with a method similar to the one described in Ref.~\cite{meyer_hydrocarbon_2008}.
    A Quantifoil grid was placed on the grown MoS$_2$ film with the amorphous carbon side of the grid facing the MoS$_2$ and a drop of isopropyl alcohol (IPA) was applied to the grid to increase adhesion between the grid and the flake.
	Potassium hydroxide (KOH) was used to etch away the SiO$_2$ layer and the samples were subsequently rinsed in deionized water and IPA for 1~min each to remove any KOH residue.
    A tabletop TEM was used to examine the sample and determine areas where a high MoS$_2$ coverage was present.
	Afterwards, these identified patches of the sample were transferred onto a Fusion AX heating chip from Protochips.
	These heating chips were calibrated for each individual chip by the company with an estimated precision of $\pm 2$~\%.
	The transfer to the heating chip was performed by placing the carbon membrane side of the TEM grid onto the chip and applying a drop of IPA to increase adhesion.
	After the IPA evaporated, the stack was placed on a heating plate for $15-20$~min at $150^\circ$C, after which the grid was ripped off with a vacuum tweezer.
	This causes parts of the amorphous carbon membrane and the attached MoS$_2$ to stick to the chip and be ripped off from the Au grid bars.
	This process was repeated until the predetermined area was successfully transferred onto the heating chip. 
	The sample was loaded into a Nion cartridge and baked overnight at a temperature of $150^\circ$C, before being introduced into the microscope magazine storage volume.
	Measurements were conducted hours to days after the transfer into the magazine storage volume was finished.

    {\bf Scanning transmission electron microscopy} 
	Measurements were carried out using a Nion UltraSTEM 100, an aberration-corrected scanning transmission electron microscope using acceleration voltages of $60$ and $90$~kV.
	The probe size of the microscope is $\sim$1~\r{A} with a beam convergence semi-angle of $30$~mrad and the base pressure inside the microscope column was below 10$^{-9}$~mbar at all times.
    The instrument is equipped with a cold field emission gun, and images were recorded using a high-angle annular dark field (HAADF) detector with a collection angle of $80-300$~mrad.
    Imaging parameters were identical to the ones used in our previous study~\cite{speckmann_combined_2023}, with a dwell time of $3$~$\mu$s/px, a flyback time of $120$~$\mu$s and images of $512~\mathrm{px}~\times~512~\mathrm{px}$.
    The resulting frame time is $0.88$~s and the time between frames was measured to be $10$~ms.
    Similar to Ref.~\cite{speckmann_combined_2023}, image series acquisition was stopped as soon as more than two missing S atoms at next-nearest neighbouring lattice sites were observed.
	Beam currents for both electron energies were measured as described in Ref.~\cite{speckmann_combined_2023}, resulting in beam currents of $106\pm 3$~pA and $196\pm 6$~pA for $60$ and $90$~keV, respectively.

	{\bf Data analysis} 
	The acquired data was analysed in the same way as described in Ref.~\cite{speckmann_combined_2023}, using a convolutional neural network (CNN) similar to the one used in Ref.~\cite{trentino_atomic-level_2021} optimizing the model created by the CNN by minimizing the intensity difference to the recorded image as described in Ref.~\cite{postl_indirect_2022}.

\section*{Acknowledgments}

This research was supported by the Austrian Science Fund (FWF) through projects [10.55776/DOC85, 10.55776/P34797] and by the Vienna Doctoral School in Physics. 
We also want to thank the group of Niall McEvoy for providing us with MoS$_2$ samples.

\bibliographystyle{elsarticle-num}
\bibliography{references}

\end{document}